# An Unpaired Cross-modality Segmentation Framework Using Data Augmentation and Hybrid Convolutional Networks for Segmenting Vestibular Schwannoma and Cochlea

Yuzhou Zhuang[1], Hong Liu[1(✉)], Enmin Song[1], Coskun Cetinkaya[2], and Chih-Cheng Hung[2]

[1] School of Computer Science and Technology, Huazhong University of Science and Technology, Wuhan, China
hl.cbib@gmail.com
[2] Center for Machine Vision and Security Research, Kennesaw State University, Marietta, MA 30060, USA

**Abstract.** The crossMoDA challenge aims to automatically segment the vestibular schwannoma (VS) tumor and cochlea regions of unlabeled high-resolution T2 scans by leveraging labeled contrast-enhanced T1 scans. The 2022 edition extends the segmentation task by including multi-institutional scans. In this work, we proposed an unpaired cross-modality segmentation framework using data augmentation and hybrid convolutional networks. Considering heterogeneous distributions and various image sizes for multi-institutional scans, we apply the min-max normalization for scaling the intensities of all scans between -1 and 1, and use the voxel size resampling and center cropping to obtain fixed-size sub-volumes for training. We adopt two data augmentation methods for effectively learning the semantic information and generating realistic target domain scans: generative and online data augmentation. For generative data augmentation, we use CUT and CycleGAN to generate two groups of realistic T2 volumes with different details and appearances for supervised segmentation training. For online data augmentation, we design a random tumor signal reducing method for simulating the heterogeneity of VS tumor signals. Furthermore, we utilize an advanced hybrid convolutional network with multi-dimensional convolutions to adaptively learn sparse inter-slice information and dense intra-slice information for accurate volumetric segmentation of VS tumor and cochlea regions in anisotropic scans. On the crossMoDA2022 validation dataset, our method produces promising results and achieves the mean DSC values of 72.47% and 76.48% and ASSD values of 3.42 mm and 0.53 mm for VS tumor and cochlea regions, respectively.

**Keywords:** Unpaired Cross-modality Segmentation, Data Augmentation, Hybrid Convolutional Networks, Vestibular Schwannoma, Cochlea.

# 1    Introduction

Unpaired cross-modality medical image segmentation aims to use labeled source scans to segment unpaired target scans without annotations between different modalities [1]. Because of the domain shift caused by different modality imaging techniques, supervised segmentation methods exhibit significant performance degradation in cross-modality segmentation [2]. Recently, unsupervised domain adaptation (UDA) methods based on adversarial learning (e.g., CycleGAN [7] and CUT [27]) are widely employed in unpaired cross-domain image processing [8], which can transfer knowledge between unpaired source and target images in an unsupervised manner. Vestibular schwannoma (VS) is a benign brain tumor originating from the vestibulocochlear nerve, and accurate segmentation of VS tumor and cochlea regions from MRI is desirable for growth detection and treatment planning of the tumor [3],[4]. The crossMoDA challenge [1] aims to automatically segment the VS tumor and cochlea regions of unlabeled high-resolution T2 scans by leveraging labeled contrast-enhanced T1 scans, and the crossMoDA2021 provided the first large cross-modality segmentation benchmark dataset with unpaired and anisotropic MR scans. Recently, the crossMoDA2022 challenge extended the segmentation task by adding multi-institutional scans. In this work, by the combination of two different image synthesis methods and an advanced hybrid convolutional network [6], we proposed an effective unpaired cross-modality segmentation framework with image alignment and multi-dimensional representations for automatically segmenting VS and cochlea regions. Moreover, to improve the robustness of our proposed method, we used various augmentation strategies to simulate the heterogeneity of VS tumor signals and overcome the differences in multi-institutional scans. Experimental results show that our proposed method produces promising results and achieves the mean Dice similarity coefficient (DSC) values of 72.47% and 76.48% and Average symmetric surface distance (ASSD) values of 3.42 and 0.53 mm for VS tumor and cochlea regions on the validation dataset of the crossMoDA2022 challenge, respectively. Our proposed method ranked 9th in the final announced evaluation results of the crossMoDA2022 challenge.

# 2    Related Works

To automatically perform VS and cochlea segmentation on hrT2 scans, the crossMoDA challenge [1] provides the benchmark datasets [3] and focuses on studying unpaired cross-modality segmentation technologies. Recent works mainly adopt the combinations of domain adaptation methods and segmentation networks for constructing unpaired cross-modality frameworks, which contain the following technologies: image alignment [13], [9]-[11], feature alignment [14], [15], disentangled representation [16] and output space alignment [17], [18]. By generating realistic images for aligning the intensity distribution between source and target domains, SynSeg-Net [13] integrated the cycle-consistent adversarial synthesis and segmentation into an end-to-end framework for cross-modality segmentation. A synergistic image and feature alignment (SIFA) framework [15] adopted the synergistic fusion of alignments from both image and feature perspectives to achieve bidirectional domain adaptation between CT and MR images. However, the standard cycle-consistency constraint assumes that the relationship between the two domains is a strong bijection, which is less effective for unpaired medical images with inconsistent anatomical structure information. To retain the anatomical structure consistency, Liu et al. [19] proposed a bidirectional domain adaptation network with anatomical structure preservation for unpaired cross-modality VS segmentation. Although the above methods have achieved impressive results in unpaired cross-modality segmentation, these methods ignored the abundant spatial information [20] and anisotropic resolutions in volumetric VS tumor brain images. On the other hand, it is hard to simulate the heterogeneity of the VS tumor signal for the methods using only a single image synthesis method [11], [12]. Thus, we integrated two different image synthesis methods with an advanced hybrid convolutional network for constructing a robust and effective cross-modality segmentation framework.

## 3  Proposed Methods

### 3.1  Overall Framework

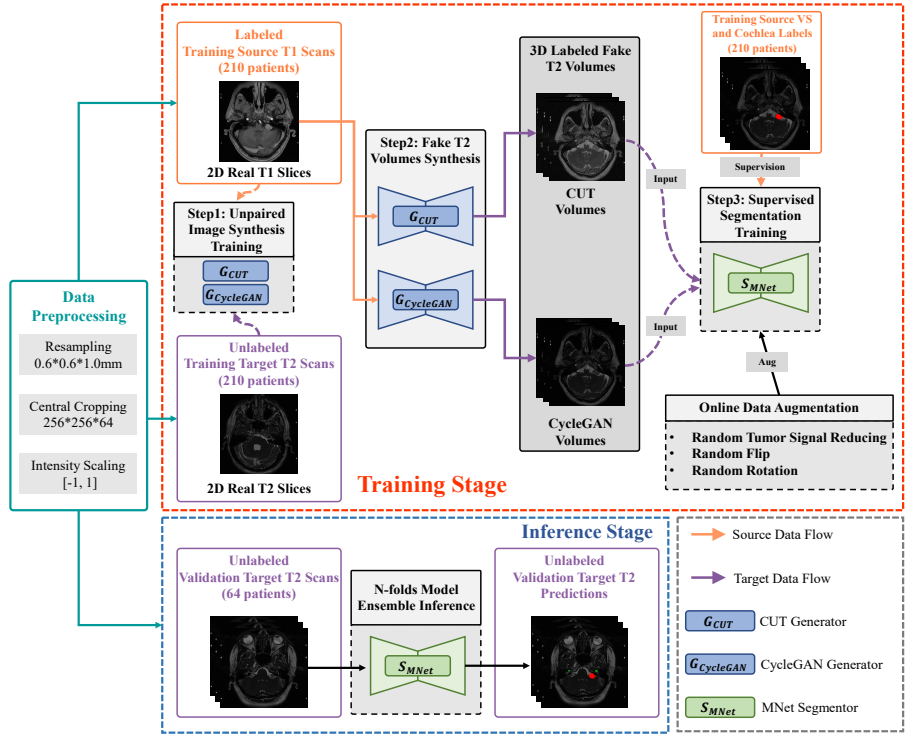

**Fig. 1.** Overall framework of our proposed method.

We proposed an unpaired cross-modality segmentation framework using various data augmentation and hybrid convolutional networks for segmenting vestibular schwannoma and cochlea, and Fig. 1 displays our proposed framework. Considering heterogeneous distributions and various image sizes for multi-institutional scans, we apply the min-max normalization for scaling the intensities of all scans and use the voxel size resampling and center cropping to obtain fixed-size sub-volumes for training. In the training stage, we first apply the UDA architectures of CUT and CycleGAN for training unpaired image synthesis models between T1 and T2 scans. Then, we use two image generators from CUT and CycleGAN to generate two groups of fake T2 volumes with different details and distributions as training data of the target domain. Finally, considering the anisotropic resolution and various image sizes of the crossMoDA 2022 dataset, we use MNet [6] as the hybrid convolutional segmentation network to segment the VS tumor and cochlea regions, which can represent sparse inter-slice information and dense intra-slice information adaptively and automatically. Besides, to improve the segmentation performance of T2 scans during training, we design a random tumor signal reducing method for simulating the heterogeneity of VS tumor signals, and combine it with random rotation and flipping for online data augmentation. By performing the 10-fold cross-validation on the crossMoDA 2022 training dataset, we can obtain 10 different trained weights for multi-model ensembles. In the inference phase, we apply the n-fold model ensembling and a sliding window with an overlap of 16 to predict unlabeled T2 scans. To reduce the time consumption of the inference phase, we choose 4-fold models to predict the testing data in the evaluation period.

### 3.2    Cross-modality Image Synthesis

To automatically and accurately segment VS and cochlea on hrT2 images, annotated ceT1 images should be converted to hrT2 images by unpaired cross-modality image synthesis methods, thereby training the segmentation model with the generated hrT2 images afterward [9], [10]. We employ the CUT and CycleGAN as image synthesis methods, and their image generators can generate realistic T2 volumes with different details and distributions for training.

Based on the official configurations of CUT [27] and CycleGAN [7], the hyperparameters of loss functions were set as the same values in the original cycle-consistency loss and patch-based contrastive loss. The ResNet with 9 residual blocks was employed as the architecture of generators, and the PatchGAN [7] was used as the adversarial discriminators. In the training stage, we train CUT and CycleGAN models independently for 100 epochs by the training batches of size $2\times256\times256\times1$. The Adam optimizer with an initial learning rate of 0.0002 and a weight decay of 0, and the number of epochs for decaying learning rate is set to 25 for both CUT and CycleGAN.

### 3.3 Online Data Augmentation

To reduce the potential overfitting risks and improve the segmentation performance in multi-institutional and anisotropic scans, we employed the following online data augmentation (ODA) methods:

- **Random Tumor Signal Reducing**. Based on [11], to introduce heterogeneity of tumor signals to mimic such clinical characteristics, we randomly reduce the image intensity of VS tumors between 0% to 50% in synthesized fake T2 scans, which is formulated as:

$$\begin{cases} I_v^{aug} = (1 - 0.5 \cdot \alpha) \cdot I_v^{ori}, v \in S_{VS} \\ I_v^{aug} = I_v^{ori}, v \notin S_{VS} \end{cases} \tag{1}$$

where $v$ is a voxel point in the input scan, $S_{VS}$ represents the voxel set of the VS tumor regions, $\alpha$ is a random factor between 0 and 1, and $I_v^{ori}$ and $I_v^{aug}$ represent the original intensity and the augmented intensity of the voxel point $v$, respectively.

- **Random Flipping.** Each of three planes is flipped randomly and independently.
- **Random Rotation.** Three different rotation angles of $90°$, $180°$, $270°$ on the axial plane are applied.

### 3.4 The Hybrid Convolutional Network for Volumetric Image Segmentation

Due to the limitations of using vanilla 3D or 2D CNNs for the segmentation of anisotropic volumes, 2.5D segmentation networks [4],[5] perform 2D convolution ($3\times3\times1$) and pooling ($2\times2\times1$) until the spacing of x/y-axis is increased to a similar level of z-axis before performing 3D convolution, thus roughly achieving balanced representation in the 3D field. However, once the spacing ratio of each axis changes, these networks must be adjusted manually and retrained to adapt images. Hence, to overcome the anisotropic resolution and various image sizes of the crossMoDA2022 dataset, as shown in Fig. 2, we use MNet [6] as the segmentation network for accurate volumetric VS and cochlea segmentation. Instead of manually adjusting the network settings according to the spacing ratio before training, MNet adaptively balances the representation inter axes in the learning process, owing to its free latent extraction and fusion of multi-dimensional representations. By sampling the training batches of size $1\times1\times256\times256\times64$, we employ the Adam optimizer with an initial learning rate of 0.0002 and a weight decay of 0.0005 to iteratively train segmentation networks for 250 epochs.

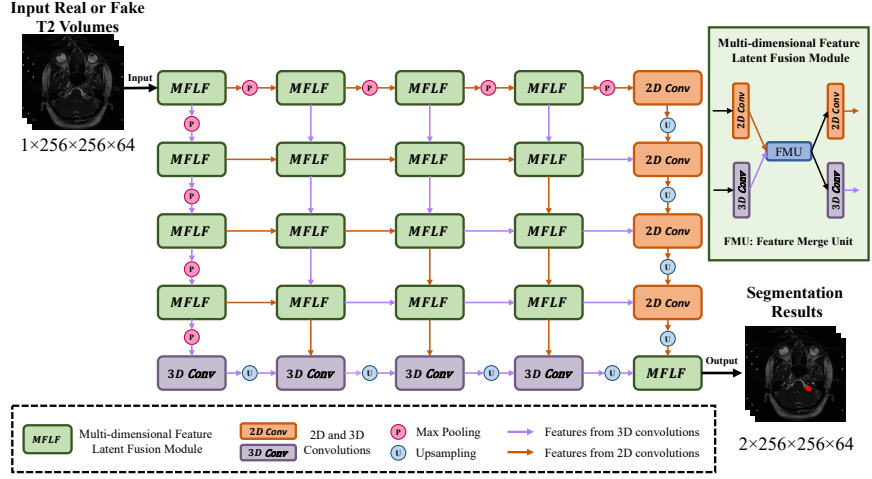

**Fig. 2**. Network architecture of MNet from the work [6], which simultaneously fuses multi-dimensional and multi-level features by adaptively learning 2D and 3D representations from anisotropic MR brain images. In MNet, 2D and 3D features are passed to the multi-dimensional feature latent fusion module which contains the combination of 2D and 3D convolutions and the feature merging unit (FMU) with element-wise subtraction.

## 4 Results

### 4.1 Implement Details

Our proposed methods were implemented using the Pytorch framework, and we performed all experiments on an Ubuntu 18.04 workstation with two 24G NVIDIA GeForce RTX 3090 GPUs and an Intel Xeon Gold 5117 2.00 GHz CPU. The crossMoDA 2022 challenge provides a training dataset with 210 labeled T1 scans and 210 unlabeled T2 scans, a validation dataset with 64 unlabeled T2 scans, and an unseen testing dataset with unlabeled T2 scans for final ranking. Due to different data sources and imaging parameters, this dataset has heterogeneous distributions and various image sizes, and it has the intra-slice spacing of (0.4~0.8) mm×(0.4~0.8) mm and the inter-slice spacing of (1.0~1.5) mm. In order to obtain the training and testing data with the same voxel size, we resample all cases to a common voxel size of 0.6×0.6×1.0 mm. Then, we utilize the min-max normalization with a maximum value of 5000 and the minimum value of 0 to scale the intensities of all scans between -1 and 1. To avoid the influence of irrelevant regions such as the eye and the skull, we use a center cropping window of size 256×256×64 (corresponding to widths, heights, and slices) to sample sub-volumes of T1 and T2 scans for training.

### 4.2 Quantitative Results

For quantitative comparison on the crossMoDA2022 dataset, the Dice Similarity Coefficient (DSC) and the Average Symmetric Surface Distances (ASSD) are used to measure the region overlap and boundary distance between the segmentation results and the ground truths, respectively. Meanwhile, the mean DSC values of VS and cochlea regions are used to compare the overall performance of different methods on crossMoDA2022 validation leaderboard. In our experiments, Table 1 shows the quantitative results of different methods on the crossMoDA2022 validation leaderboard, where 'ODA' denotes the online data augmentation, and 'n-folds' denotes the n-fold model ensembles. As shown in Table 1, Method #1 without domain adaptation showed the lowest segmentation performance because of the domain shift problem between the T1 and T2 modalities.

By employing CycleGAN to generate realistic T2 images for unsupervised domain adaptation, Method #2 significantly improves the mean DSC values from 58.69% to 61.76% compared to Method #1. However, the 2D UNet of Method #2 cannot obtain rich space information in segmentation processes, which causes the problem that the segmentation performance of Method #2 is lower than Method #3 and #4 using 3D convolutions. By using MNet to adaptively fuse 2D and 3D representations, Method #5 without model ensembles obtains better overall performance of VS and cochlea regions than Method #3 and #4 with 5-fold model ensembles. By adding generative and online data augmentation, our proposed method (Method #7) gains the best overall DSC value of 74.48% in our experiments, which achieves the DSC values of 72.47% and 76.48% and ASSD values of 3.42 and 0.53 mm for VS tumor and cochlea on the validation dataset, respectively.

**Table 1.** Quantitative results of different methods on crossMoDA2022 validation leaderboard, where 'ODA' denotes the online data augmentation and 'n-folds' denotes the n-fold model ensembles. Bold values indicate the best results.

| # | Methods | | DSC (%)↑ | | | ASSD (mm)↓ | |
|---|---|---|---|---|---|---|---|
| | Synthesis | Segment | VS | Cochlea | Mean | VS | Cochlea |
| 1 | None | UNet2D (5folds) | 36.75±29.30 | 1.3±2.66 | 19.05±14.58 | 9.04±12.19 | 13.74±10.65 |
| 2 | CycleGAN | UNet2D (5folds) | 65.88±22.27 | 69.42±6.29 | 67.65±12.72 | 4.84±5.24 | 0.62±1.78 |
| 3 | CycleGAN | UNet3D (5folds) | 64.13±93.53 | 72.92±4.82 | 68.52±12.58 | 6.17±7.78 | 0.59±1.78 |
| 4 | CycleGAN | UNet2.5D (5folds) | 64.74±27.09 | 73.33±4.71 | 69.03±14.25 | **2.44±3.21** | 0.37±0.16 |
| 5 | CycleGAN | MNet (1fold) | 68.56±21.48 | 72.23±4.68 | 70.39±11.29 | 3.63±6.91 | 0.59±1.77 |
| 6 | CycleGAN+CUT | MNet (ODA+4folds) | 68.98±23.56 | 74.14±4.87 | 71.56±12.57 | 2.91±5.88 | **0.36±0.17** |
| 7 | CycleGAN+CUT | MNet (ODA+10folds) | **72.47±17.94** | **76.48±4.71** | **74.48±9.99** | 3.42±3.53 | 0.53±1.78 |

## 4.3 Qualitative Results

Fig. 3 visualizes the segmentation results of different methods in Table 1 for qualitative analysis. As shown in Fig. 3, due to the significant domain gap between training T1 modality and testing T2 modality, Method #1 without domain adaptation hardly segments the cochlea regions, and the segmentation results of the VS region are also incomplete. By generating realistic target domain images for alleviating the intensity distribution gap, Method #2, #3, and #4 obtain more accurate segmentation results than Method #1, especially for the cochlea regions. Unlike the above methods, our proposed Method #6 further aligns the intensity distribution of VS and cochlea regions by various augmentation methods and adaptively obtains multi-dimensional representations, which obtains stable and accurate segmentation results. Finally, our proposed framework (Method #7) with 10-fold model ensembles effectively alleviates the performance degradation of cross-modality VS and cochlea segmentation in Fig. 3.

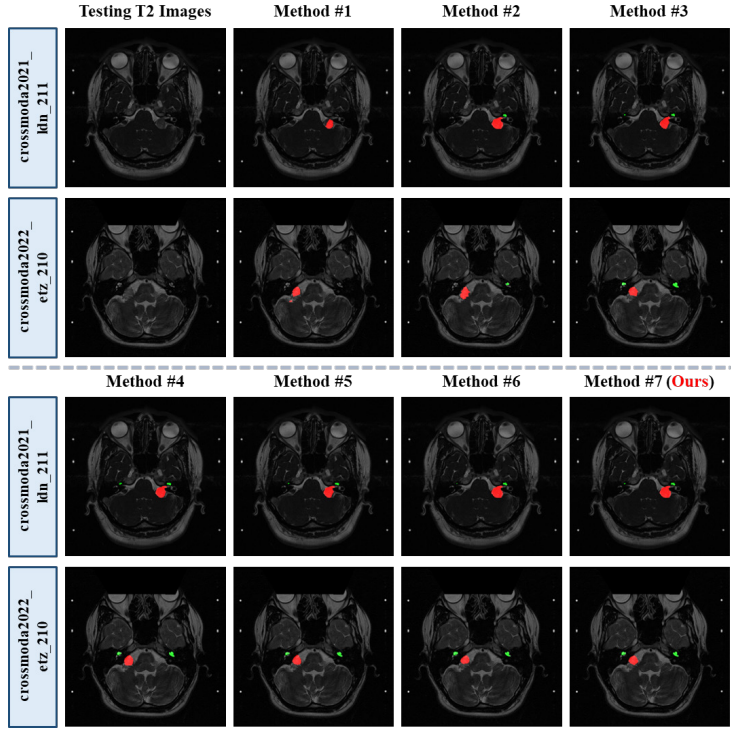

**Fig. 3.** Qualitative comparison of segmentation results of different methods on the crossMoDA2022 validation dataset, where the VS and cochlea regions o correspond to red and green colors, respectively.

## 5    Discussion

In this section, we discuss the advantages and weaknesses of our proposed framework from two main perspectives: 1) Cross-modality Image Synthesis Method, and 2) Segmentation Networks.

### 5.1    Cross-modality Image Synthesis Methods

For cross-modality VS and cochlea segmentation in the crossMoDA2022 challenge, current UDA methods (e.g., CycleGAN [7] and CUT [27]) can effectively help segmentation networks to learn the intensity distribution and annotation knowledge of unlabeled target domain by generative adversarial learning between T1 and T2 modalities. However, due to the significant heterogeneity of the VS and cochlea regions in cross-modality multi-institutional scans, it is difficult for a single image synthesis method to overcome the distribution gap among different modalities or scans, resulting in sub-optimal performance. As shown in Fig. 4, we simultaneously employ both CycleGAN and CUT to generate various and realistic target T2 images from labeled T1 images for improving the generalization performance and stability of the segmentation network. On the other hand, in Fig. 4, we observe that the synthesis results of CUT and CycleGAN are different but complementary, which is beneficial for generative data augmentation. The shortcoming of our proposed method is that two individual generative networks need to be trained, which is less efficient compared to other end-to-end methods.

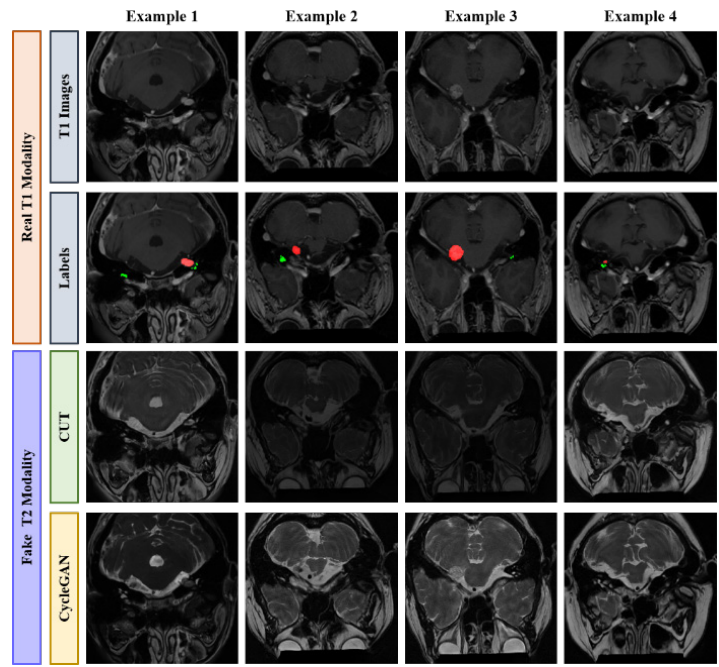

**Fig. 4.** Comparison of cross-modality image synthesis results for two different image synthesis methods on the crossMoDA2022 training dataset.

### 5.2 Segmentation Networks

Unlike the existing methods [9]-[12] using nnU-Net, 3D UNet, or 2.D UNet, we used the MNet with the appropriate number of parameters and better multi-dimensional representations as the segmentation network. Compared with other segmentation networks, the method using MNet [6] achieves a better mean DSC value in Table 1. Besides, compared to UNet3D with 22.58M and nnU-Net with 28.50M, MNet with 8.7M parameters shows great superiority in terms of the number of parameters, which means lower model complexity in the crossMoDA2022 challenge.

## 6    Conclusions

In this work, we proposed an unpaired cross-modality domain adaptation framework for VS and cochlea segmentation. Our proposed framework consists of synthesis and segmentation components. We applied two different image synthesis methods and various data augmentations to deal with the MRIs from different sites and scanners, and we utilized an effective hybrid convolutional network to adaptively integrate sparse inter-slice information with dense intra-slice formation for accurate segmentation. In the validation stage of crossMoDA2022, our method shows promising results and achieves the DSC values of 72.47% and 76.48% and ASSD values of 3.42 and 0.53 mm for VS tumor and cochlea on the validation dataset, respectively.

## References

[1]  R. Dorent et al., "CrossMoDA 2021 challenge: Benchmark of cross-modality domain adaptation techniques for vestibular schwannoma and cochlea segmentation," Med. Image Anal., p. 102628, 2022, doi: 10.1016/j.media.2022.102628..

[2]  R. Dorent et al., "Scribble-based Domain Adaptation via Co-segmentation," in International Conference on Medical Image Computing and Computer-Assisted Intervention, 2020, pp. 479–489

[3]  J. Shapey et al., "Segmentation of vestibular schwannoma from MRI, an open annotated dataset and baseline algorithm," Sci. Data, vol. 8, no. 1, p. 286, 2021, doi: 10.1038/s41597-021-01064-w.

[4]  J. Shapey et al., "An artificial intelligence framework for automatic segmentation and volumetry of vestibular schwannomas from contrast-enhanced T1-weighted and high-resolution T2-weighted MRI," J. Neurosurg., vol. 134, no. 1, pp. 171–179, 2019.

[5]  G. Wang et al., "Automatic segmentation of vestibular schwannoma from T2-weighted mri by deep spatial attention with hardness-weighted loss," in Proc. Int. Conf. Med. Image Comput. Comput.-Assisted Intervention, China: Springer, 2019, pp. 264–272.

[6]  Z. Dong et al., "MNet: Rethinking 2D/3D Networks for Anisotropic Medical Image Segmentation," 2022, [Online]. Available: http://arxiv.org/abs/2205.04846.

[7]  J.-Y. Zhu et al., "Unpaired image-to-image translation using cycle-consistent adversarial networks," in Proceedings of the IEEE international conference on computer vision, 2017, pp. 2223–2232

[8]  T. Park et al., "Contrastive learning for unpaired image-to-image translation," in Proceedings of the European conference on computer vision (ECCV), 2020, pp. 319–345

[9] H. Shin, H. Kim, S. Kim, Y. Jun, T. Eo, and D. Hwang, "COSMOS: Cross-Modality Unsupervised Domain Adaptation for 3D Medical Image Segmentation based on Target-aware Domain Translation and Iterative Self-Training," arXiv Prepr. arXiv2203.16557, 2022.

[10] H. Dong, F. Yu, J. Zhao, B. Dong, and L. Zhang, "Unsupervised Domain Adaptation in Semantic Segmentation Based on Pixel Alignment and Self-Training," pp. 4–8, 2021, [Online]. Available: http://arxiv.org/abs/2109.14219.

[11] J. W. Choi, "Using Out-of-the-Box Frameworks for Unpaired Image Translation and Image Segmentation for the crossMoDA Challenge," pp. 1–5, 2021, [Online]. Available: http://arxiv.org/abs/2110.01607.

[12] H. Liu, Y. Fan, C. Cui, D. Su, A. McNeil, and B. M. Dawant, "Unsupervised Domain Adaptation for Vestibular Schwannoma and Cochlea Segmentation via Semi-supervised Learning and Label Fusion," vol. 1, pp. 1–11, 2022, [Online]. Available: http://arxiv.org/abs/2201.10647.

[13] Y. Huo et al., "Synseg-net: Synthetic segmentation without target modality ground truth," IEEE Trans. Med. Imaging, vol. 38, no. 4, pp. 1016–1025, 2018.

[14] Q. Dou et al., "PnP-AdaNet: Plug-and-play adversarial domain adaptation network at unpaired cross-modality cardiac segmentation," IEEE Access, vol. 7, pp. 99065–99076, 2019

[15] C. Chen et al., "Unsupervised bidirectional cross-modality adaptation via deeply synergistic image and feature alignment for medical image segmentation," IEEE Trans. Med. Imaging, vol. 39, no. 7, pp. 2494–2505, 2020.

[16] C. Pei, F. Wu, L. Huang, and X. Zhuang, "Disentangle domain features for cross-modality cardiac image segmentation," Med. Image Anal., vol. 71, p. 102078, 2021.

[17] Y.-H. Tsai et al., "Learning to adapt structured output space for semantic segmentation," in Proceedings of the IEEE conference on computer vision and pattern recognition, 2018, pp. 7472–7481.

[18] S. Vesal et al., "Adapt Everywhere: Unsupervised Adaptation of Point-Clouds and Entropy Minimization for Multi-Modal Cardiac Image Segmentation," IEEE Trans. Med. Imaging, vol. 40, no. 7, pp. 1838–1851, 2021.

[19] H. Liu et al., "A bidirectional multilayer contrastive adaptation network with anatomical structure preservation for unpaired cross-modality medical image segmentation," Comput. Biol. Med., p. 105964, 2022.

[20] K. Yao et al., "A novel 3D unsupervised domain adaptation framework for cross-modality medical image segmentation," IEEE J. Biomed. Heal. Informatics, p. 1, 2022.